\documentclass[aps,prb,twocolumn,showpacs,amsmath,amssymb,superscriptaddress,floatfix]{revtex4-2}
\usepackage{graphicx}
\usepackage{CJK}
\usepackage{color}
\usepackage[colorlinks,bookmarks=false,citecolor=blue,linkcolor=red,urlcolor=blue]{hyperref}
\usepackage{multirow}
\usepackage{ulem}
\usepackage{tabularx}
\usepackage[nounderscore]{syntax}
\graphicspath{{figs/}}

\newcommand{\be}{\begin{equation}}
\newcommand{\ee}{\end{equation}}

\begin{document}
\title{Surface phase transitions in a (1+1)-dimensional $SU(2)_1$ conformal field theory boundary coupled to a (2+1)-dimensional $Z_2$ bulk}

\author{Zhe Wang}
\affiliation{Department of Physics, School of Science and Research Center for Industries of the Future, Westlake University, Hangzhou 310030,  China}
\affiliation{Institute of Natural Sciences, Westlake Institute for Advanced Study, Hangzhou 310024, China}

\author{Shang-Qiang Ning}
\affiliation{Department of Physics, The Chinese University of Hong Kong, Sha Tin, New Territory, Hong Kong, China}

\author{Zenan Liu}
\affiliation{State Key Laboratory of Optoelectronic Materials and Technologies, Guangdong Provincial Key Laboratory of Magnetoelectric Physics and Devices, Center for Neutron Science and Technology, School of Physics, Sun Yat-Sen University, Guangzhou 510275, China}

\author{Junchen Rong}
\affiliation{Institut des Hautes \'Etudes Scientifiques, 91440 Bures-sur-Yvette, France}

\author{Yan-Cheng Wang}
\email{ycwangphys@buaa.edu.cn}
\affiliation{Hangzhou International Innovation Institute, Beihang University, Hangzhou 311115, China}
\affiliation{Tianmushan Laboratory, Hangzhou 311115, China}

\author{Zheng Yan}
\email{zhengyan@westlake.edu.cn}
\affiliation{Department of Physics, School of Science and Research Center for Industries of the Future, Westlake University, Hangzhou 310030,  China}
\affiliation{Institute of Natural Sciences, Westlake Institute for Advanced Study, Hangzhou 310024, China}

\author{Wenan Guo}
\email{waguo@bnu.edu.cn}
\affiliation{School of Physics and Astronomy, Beijing Normal University, Beijing 100875, China}
\affiliation{Key Laboratory of Multiscale Spin Physics (Ministry of Education), Beijing Normal University, Beijing 100875, China}
\date{\today}
\begin{abstract}
We design a (2+1))-dimensional [(2+1)D] quantum spin model in which spin-1/2 ladders are coupled through antiferromagnetic Ising interactions.  
The model hosts a quantum phase transition in the (2+1)D $Z_2$ universality class from the Haldane phase to the antiferromagnetic Ising ordered phase. We focus on studying the surface properties of three different surface configurations when the Ising couplings are tuned. Different behaviors are found on different surfaces.
We find ordinary and two different extraordinary surface critical behaviors (SCBs) at the bulk critical point.
The ordinary SCBs belong to the surface universality class of the classical 3D Ising bulk transition. 
One extraordinary SCBs is induced by the topological properties of the Haldane phase.
Another extraordinary SCBs at the bulk critical point is induced by an unconventional surface phase transition where the surface develops an Ising order before the bulk. 
This surface transition is realized by coupling a (1+1)-dimensional [(1+1)D] $SU(2)_1$ CFT boundary to a (2+1)D bulk with $Z_2$ symmetry. 
We find that the transition is neither a (1+1)D $Z_2$ transition, expected based on symmetry consideration,
nor a Kosterlitz-Thouless-like transition, violating the previous theoretical prediction. This new surface phase transition and related extraordinary SCBs deserve further analytical and numerical exploration.
\end{abstract}
\maketitle


\section{Introduction}
	\label{intro}
A $d$-dimensional(D) system with $(d-1)$ D surfaces/boundaries can have rich critical behaviors. Generally speaking, when the bulk reaches the disorder-order critical point from the disordered phase, the surface remains disordered but shows critical singularity induced by the bulk criticality. This is called ``ordinary surface critical behaviors (SCB)''. However, if the surface couplings are sufficiently enhanced, the surface may order while the bulk stays disordered if the dimensions of the surface are larger than the lower critical dimensions $d_l$ of the system. This is a surface transition in the ($d-1$)D system with short-range interactions. The surface exhibits extra singularities at the bulk critical point, called the ``extraordinary SCBs". As the surface coupling enhancement is reduced, the surface transition and the bulk transition meet at a multi-critical point where ``special SCBs" show up~\cite{Cardy, Binder1974, Binder1983, Deng2005}. More recently, a new class of SCBs at a bulk critical point, the ``extraordinary-log SCBs", is revealed for 3D systems with continuous symmetry but surface dimensions smaller than $d_l$ \cite{Metlitski2020, ParisenToldin2021, Hu2021, ParisenToldin2022, meineri2022, Zou2022}. 

Classically, these rich critical behaviors are attributed to the fact that the local environment of a given degree of freedom near the surface is different from that deep inside the bulk~\cite{Cardy}. This general picture of SCBs should also apply to a quantum critical point (QCP), based on the mapping between a $d$-dimensional quantum system and a $(d + 1)$D classical system.
However, when the quantum mechanism intervenes, exotic surface states with purely quantum mechanic origin arise.
The surface of the system can be gapless in a symmetry-protected topological (SPT) phase \cite{Haldane, AKLT, XiaoGang2009, Frank2010, XiaoGang2012},
the boundary formed by dangling spin-1/2 chain of a topological trivial gapped phase is gapless due to the topological $\theta$ term
\cite{Zhang2017, Ding2018, Weber2018, Weber2019, Wang2022}. 
When these gapless surface modes are coupled to the bulk critical mode at the (2+1)D O(3) bulk critical point, exotic nonordinary multicritical  SCBs \cite{Zhang2017, Zhu2021, Wang2022} or extraordinary SCB \cite{Wang2023} are present. 

One way to understand these exotic quantum SCBs is that the bulk critical 
fluctuations yield effective nonlocal interactions in space-time, instead of instantaneous nonlocal 
interactions, at the boundary, which makes the (1+1)D $SU(2)_1$ boundary conformal field theory (CFT) 
unstable~\cite{Jian2021}.
In particular, when a (1+1)D $SU(2)_1$ CFT boundary is coupled to the (2+1)D Ising bulk critical fluctuations, 
it is demonstrated through renormalization group calculations that, 
the surface goes through a transition from a gapless $SU(2)_1$ CFT to an Ising ordered phase before the bulk actually hits criticality \cite{Jian2021}.
This surface transition is argued to be in the Kosterlitz-Thouless-like (KT-like) 
universality class (UC), akin to the transition from a Luttinger liquid to a 
$Z_2$ valence bond solid (VBS) phase in a purely 1D spin-1/2 chain with both
nearest and next-nearest neighbor Heisenberg interactions \cite{Haldane1982}, rather than a (1+1)D Ising transition.

\begin{figure}[htb]
\centering
\includegraphics[width=0.49\textwidth]{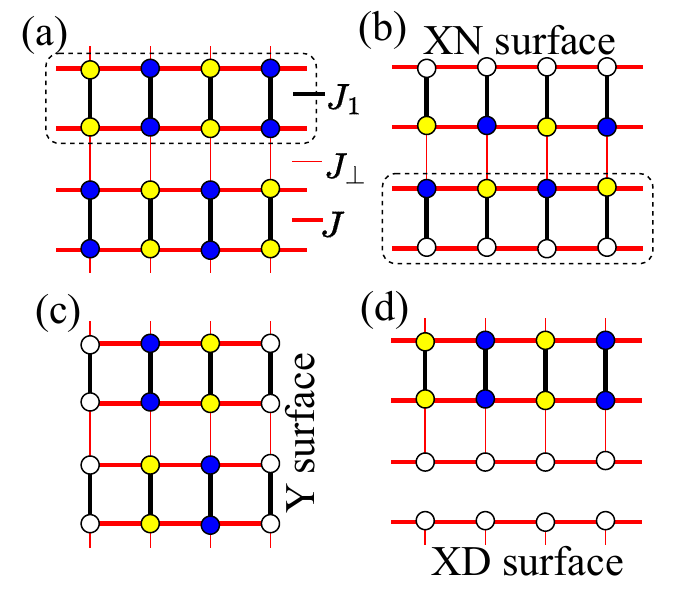}
\caption{ The two-dimensional coupled usual ladders. The lattice is bipartite with sublattices A (yellow circles) and B (blue circles). (a) Periodic boundary conditions(PBCs) are applied in the $x$ and $y$ directions. A usual ladder is shown inside the dashed rectangular box. The Ising interladder couplings $J_{\perp}>0$ are indicated by thin red lines, the intraladder Heisenberg couplings $J_{1}<0$ by thick black lines, and the intraladder Heisenberg couplings $J>0$ by thick red lines. (b) PBCs are used in the $x$ direction, while the weak bonds $J_{\perp}$ are cut to expose the XN surface. (c) PBCs are used in the $y$ direction, while open boundary conditions are applied in the $x$ direction to expose the Y surface (d) Periodic boundary conditions are used in $x$ direction, while cut the strong bonds $J_{1}$ to expose the XD surface. Open circles denote spins on the surfaces. }
	\label{Fig:model}
\end{figure}

In this paper, we study the rich universal surface physics of a quantum spin system using quantum Monte Carlo (QMC) simulations. We design a (2+1)D quantum spin model (see Fig.~\ref{Fig:model}) hosting a quantum phase transition  
in the (2+1)D $Z_2$ universality class from a topological nontrivial Haldane phase to an antiferromagnetic (AF) Ising phase. Three different surface configurations, as illustrated in Fig.~\ref{Fig:model}(b)-(d), are investigated. We find different surface behaviors on different surfaces. For the surface with gapped surface states in the bulk Haldane phase, we find ordinary SCBs of the 3D Ising UC at the bulk critical point. For the surface, which is ordered throughout the bulk Haldane phase as a result of the SPT properties, we find the extraordinary SCBs as expected.
Most interestingly, for the surface formed by dangling spins, which is a (1+1)D $SU(2)_1$ CFT, we find that the surface orders before the bulk as the Ising couplings increase, which is consistent with the theoretical prediction. However, our numerical results find, violating the prediction, that this transition is neither a (1 + 1)D Ising transition nor a KT-like transition, pointing to a new universality class. In addition, at the bulk critical point, the surface undergoes an unconventional extraordinary SCB, with exponents different from the extraordinary SCBs on the other surface. 
    
\section{Models and methods}
\label{Sec:mm}
We study the spin-1/2 Heisenberg model on a designed 2D bipartite lattice 
constructed by coupling usual ladders\cite{diagonal2000} with AF Ising interactions, see Fig. \ref{Fig:model}. 
We will refer to the lattice as Ising coupled usual ladders.
The Hamiltonian is given by
 \begin{equation}
	\begin{split}
	H &= \sum_{j=0}H_{j}+ J_{\perp}\sum_{i,j=0}S^z_{i,2j+1}\cdot S^z_{i,2(j+1)},
		\end{split}
	\end{equation}
where the first sum is over the usual ladders with $H_{j}$ describing the $j$-th ladder written as follows 
	\begin{equation}
	\begin{split}
	H_j &= J \sum_{l=0,1} \sum_i \mathbf{S}_{i,2j+l}\cdot \mathbf{S}_{i+1,2j+l}\\
	&+ J_1\sum_{i} \mathbf{S}_{i,2j}\cdot \mathbf{S}_{i,2j+1},
		\end{split}
		\label{H_ladder}
	\end{equation}
where $l=0,1$ denote two legs of the $j$-th usual ladder, $J>0$ and $J_1<0$ is the intraladder Heisenberg exchange coupling; The second sum describes the couplings of the neighboring ladders with the interladder Ising coupling $J_{\perp}>0$. 

We set $J=|J_1|=1$ to fix the energy scale. When $J_{\perp}$ is comparable to $J$, the model is expected in the AF Ising phase. For the limit $J_{\perp} \to 0$,  the model is adiabatically connected to the 1D usual ladder, which behaves like a spin-1 chain \cite{diagonal2000} and hence the model is tuned into a quasi-one-dimensional Haldane (Q1DH) phase. These two phases are separated by a (2+1)D $Z_2$ critical point revealed by our simulations described below. In this work, we use the stochastic series expansion (SSE) algorithm \cite{Sandviksusc1991,Sandvik1999,wu2023classical, yan2022global, yan2019sweeping} to explore the bulk and surface behaviors. PBCs are applied along both $x$ and $y$ directions when the bulk properties are studied. In our simulations, we have reached linear size up to $L=128$, and the inverse temperature scales as $\beta=L$. Typically $10^{8}$ MC samples are taken for each coupling strength. 

\section{ Bulk Results}
\label{sec:bulk}
The bulk transition is associated with the spontaneous breaking of the spin-flip ($Z_2$) symmetry. The staggered magnetization $m_s^z=\frac{1}{L^2}\sum_i \phi_{i} S_i^z$ is used to describe the AF Ising order with the staggered phase factor $\phi_{i}=\pm 1$ according to the sublattice. The dimensionless Binder ratio $R_{2}$ \cite{Binder1981,Binder1984} is defined using $m_s^z$ as $R_{2}=\frac{\langle (m_{s}^{z})^{4}\rangle}{\langle (m_{s}^{z})^{2}\rangle^{2}}$, and the corresponding Binder cumulant $U_{2}$  is written as $U_{2}= \frac{1}{2}\left(3-R_2\right)$. $U_{2}$ converging to 1 with increasing system size indicates the existence of magnetic order, while approaching zero with increasing system size implies that the system is in the magnetically disordered phase. 

\begin{figure}[htb]
\centering
\includegraphics[width=0.49\textwidth]{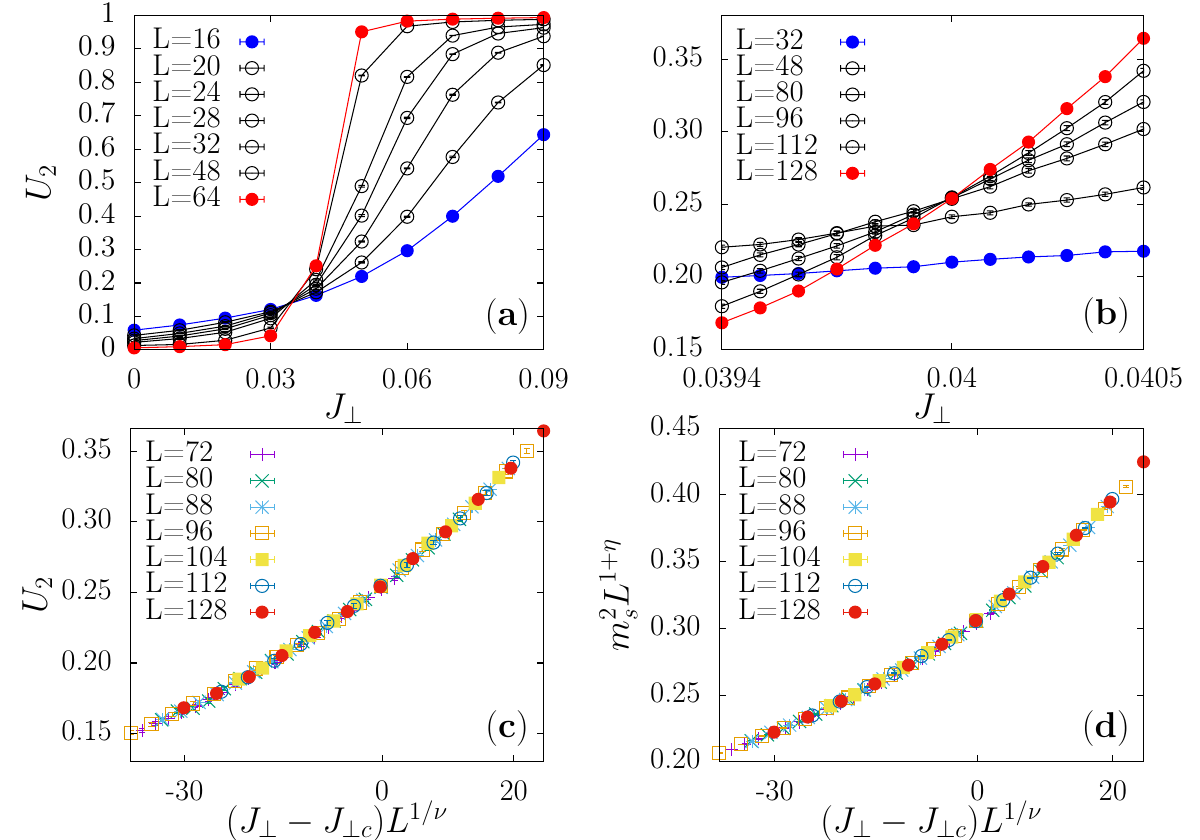}
\caption{ Critical properties of the bulk phase transition. (a) Binder cumulant $U_{2}$ versus $J_{\perp}$ for different system sizes. (b) shows details around the crossing point for large system sizes. (c) Data collapse of $U_2$ using $\nu=0.64$ and $J_{\perp c}=0.040006$. (d) Data collapse of $m_s^2$ using $\nu=0.64$, $J_{\perp c}=0.040006$, and $\eta=0.036$. Error bars are much smaller than the symbols.}
\label{fig:binder}
\end{figure}

Figure \ref{fig:binder}(a) and (b) show $U_2$ as functions of $J_{\perp}$ for different system sizes. Clearly, the model is in the AF ordered phase when $J_{\perp}$ is larger than 0.04. Since $U_2$ is dimensionless at a critical point, the crossings of curves for different sizes roughly indicate the transition point.

We adopt the finite-size scaling formula
\begin{equation}
	A(J_{\perp},L)=L^{\kappa}f[(J_{\perp}-J_{\perp c})L^{1/\nu},L^{-\omega}]
\label{scaling}
\end{equation}
to extract  the critical coupling $J_{\perp c}$ and the correlation length exponent $\nu$, where $f(x)$ is a scaling function 
with $\omega>0$ the effective correction to scaling exponent. 
For $A=U_2$, which is dimensionless,  $\kappa=0$. 
One can observe, in Fig. \ref{fig:binder} (b), that the crossings of $U_2$ curves for two neighboring sizes shifts to larger $J_\perp$ as 
sizes increase as results of the correction to scaling. 
Such a large correction to scaling can be attributed to the strong anisotropic nature of this model:
the correlation along $x$ direction is much longer than that along the $y$ direction \cite{Matsumoto2001}.
It is possible to reduce this correction to scaling by changing the aspect ratio of the systems, following \cite{Matsumoto2001,Zhu2021}.
In Appendix \ref{sec:r4}, 
we present simulation results on systems with aspect ratio $R=L_x/L_y=4$ and $\beta=4L_y$. For these systems, 
the shifts of the crossings of $U_2$ curves become much narrower.

The corrections to scaling are included in scaling formula Eq. (\ref{scaling}), which we can fit to our finite-size data 
$U_2(J_\perp, L)$. We do the fitting in an alternative way: we only use data of sufficient large sizes, i.e. $L\ge 72$, in the fitting,
for which the effect of corrections to scaling is neglectable. In such a case, 
we expand $f(x)$ to polynomials near the transition point. We obtain $J_{\perp c}=0.040006(3)$ and  $\nu=0.64(1)$ by fitting the 
polynomials to the finite-size data. 
Figure \ref{fig:binder} (c) shows a perfect data collapsing of $U_2$ using the obtained $J_{\perp c}$ and $\nu$. 

Next, the squared 
magnetization, $m_s^2=\langle (m_s^z)^2\rangle$, are analyzed to extract the exponent $\eta$. We expect $m_s^2$ scales in the form of Eq.~(\ref{scaling}) with $\kappa=-(1+\eta)$. 
$f(x)$ is another scaling function for sufficient large sizes. Using the obtained  $J_{\perp c}$ and $\nu$, we obtain 
$\eta=0.036(3)$ by expanding $f(x)$ to polynomials and fitting it to finite-size data of $m_s^2(J_\perp,L)$. 
Perfect data collapse using obtained critical properties is shown in Fig. \ref{fig:binder} (d). 

We have also performed similar finite-size scaling analysis on data from systems with $R=4$.   
The obtained critical point and exponents are consistent with those found for $R=1$ systems, 
indicating that the critical properties of the system are not affected by the spatial anisotropy. See Appendix \ref{sec:r4} for details.

Comparing with the best-known exponents of the 3D Ising UC\cite{Deng2003}, we conclude that the critical point belongs to the 3D Ising UC. This also demonstrates that the nontrivial topological property of the Haldane phase does not affect the universal properties of the bulk phase transition, as noticed in \cite{Zhu2021}.
\section{ Surface  behaviors}

\label{sec:scb}
We now investigate the surface behaviors on the XN, XD, and Y surfaces, respectively. We calculate the surface parallel correlation $C_\parallel(L/2)$ between two surface spins $i$ and $j$ with the longest distance $|i-j|=L/2$ and the perpendicular correlation $C_\perp(L/2)$ averaging $C(\mathbf{r}_{ij})$ between spin $i$ fixed on the surface and spin $j$ located at the center of the bulk, with $\mathbf{r}_{ij}$ perpendicular to the surface and $|j-i|=L/2$.

The surface staggered magnetic susceptibility $\chi_{1s}$ with respect to the surface field $h_1$ is also calculated through the Kubo formula \cite{Sandviksusc1991} 
\be
\chi_{1s}
= L\int_{0}^{\beta}d\tau\langle m^z_{1s}(\tau)m^z_{1s}(0)\rangle, 
\ee
where $m^z_{1s}$ is the staggered surface magnetization 
\be
m_{1s}^z =\frac{1}{L}\sum_{i \in {\rm surface}} \phi_i S_i^z,
\ee
where the summation is restricted on the surface, $\phi_i=\pm 1$ depending on the sublattice to which $i$ belongs. 
Based on $m_{1s}^z$, we define the surface Binder ratio as
\be
R_{12}=\frac{\langle (m_{1s}^{z})^{4}\rangle}{\langle (m_{1s}^{z})^{2}\rangle^{2}} 
\ee
and the Binder cumulant as \cite{Binder1981,Binder1984}
\be
U_{12}= \frac{1}{2}\left(3-R_{12}\right). 
\ee

At the bulk critical point, the scaling of the surface susceptibility $\chi_{1s}$ and the squared  magnetization  are represented by $y_{h1}$  
the scaling dimension of the surface field $h_1$ \cite{Binder1974} :
\begin{equation}
\chi_{1s} = a_1 L^{-(d+z-1-2y_{h1})} +a_2 L^{-1} +\cdots,
\label{suchis}
\end{equation}
and
\begin{equation}
m^2_{1s}(L) = m^2_{1s}+ b_1 L^{-2(d+z-1-y_{h1})} +b_2 L^{-1} +\cdots.
\label{sum1s}
\end{equation}
The two correlations are characterized by two anomalous dimensions $\eta_\parallel$ and $\eta_\perp$, respectively;
 \begin{equation}
C_{\parallel}(L/2) =C_\parallel+c_1 L^{-(d+z-2+\eta_{\parallel})}+c_2 L^{-1}+\cdots,
\label{sucs1}
\end{equation}
and
\begin{equation}
C_{\perp}(L/2) = d_1 L^{-(d+z-2+\eta_{\perp})}+d_2 L^{-1}+\cdots.
\label{sucs2}
\end{equation}
Here, for our model $d=2$ and $z=1$;
$a_i$, $b_i$, $c_i$ and $d_i$ are unknown constants; the $1/L$ terms are the leading correction; $C_\parallel=m^2_{1s}=0$  in ordinary or special SCBs; and $C_\parallel $ should be equal to $m^2_{1s}$ to characterize the ordered surface in an extraordinary SCB.

The scaling dimensions are related through the following scaling relations \cite{Diehl}:
 \begin{equation}
2\eta_\perp = \eta_\parallel +\eta,  ~~~~~\eta_\parallel = d+z - 2y_{h1},  
\label{scalings1} 
\end{equation}
where $\eta$ is the anomalous magnetic scaling dimension of the bulk critical point.

\subsection{Ordinary SCBs on the XN surface}

\begin{figure}[htb]
	\centering
	\includegraphics[width=0.46\textwidth]{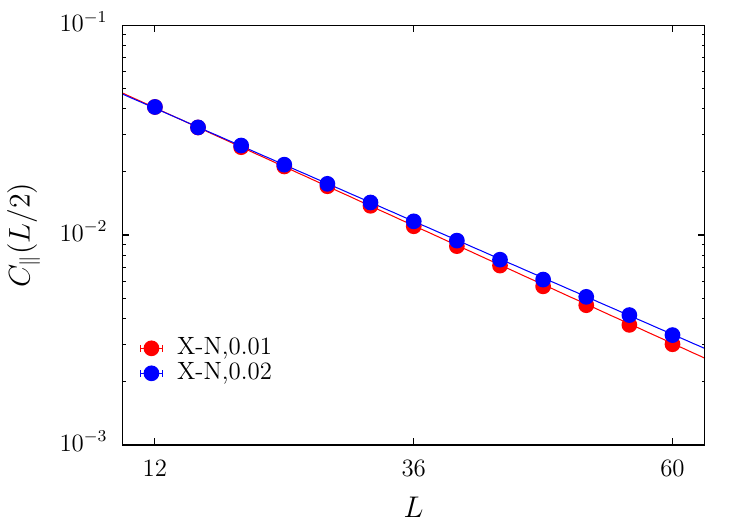}
	\caption{Surface correlation $C_{\parallel}(L/2)$ vs. system size $L$ in the Haldane phase ($J_{\perp}=0.01,0.02$). The plot is set on a linear-log scale. Exponentially decaying with $L$ is observed, meaning the surface states are gapped.}
	\label{fig:xngap}
\end{figure}

We first study the surface critical behaviors on the XN surface exposed by cutting the weak bonds $J_1$, see Fig. \ref{Fig:model} (b). 
The surface is formed by nondangling spins, which are one leg of the dangling ladder; hence, the spectrum of the surface is gapped when the bulk is in the Q1DH phase. 
To study the surface states, we calculate the surface parallel correlation $C_\parallel(L/2)$ and observe an exponential decay of $C_\parallel(L/2)$ supporting this spectrum. 
 Figure \ref{fig:xngap} shows $C_\parallel(L/2)$ at $ J_\perp=0.01$ and $ J_\perp=0.02$ sitting in the Q1DH phase. The data
can be fitted using straight lines on a linear-log scale, meaning
the correlation decays exponentially. Fitting the curves with
\begin{equation}
    C_\parallel(L/2) \sim \exp{(-L/a)},
    \label{surface_gap}
\end{equation}
we obtain $a=18.56(7)$ at $J_\perp=0.01$ and $a=19.35(8)$ at $J_\perp =0.02$, which means gapped surface state.

\begin{figure}[htb]
	\centering
	\includegraphics[width=0.46\textwidth]{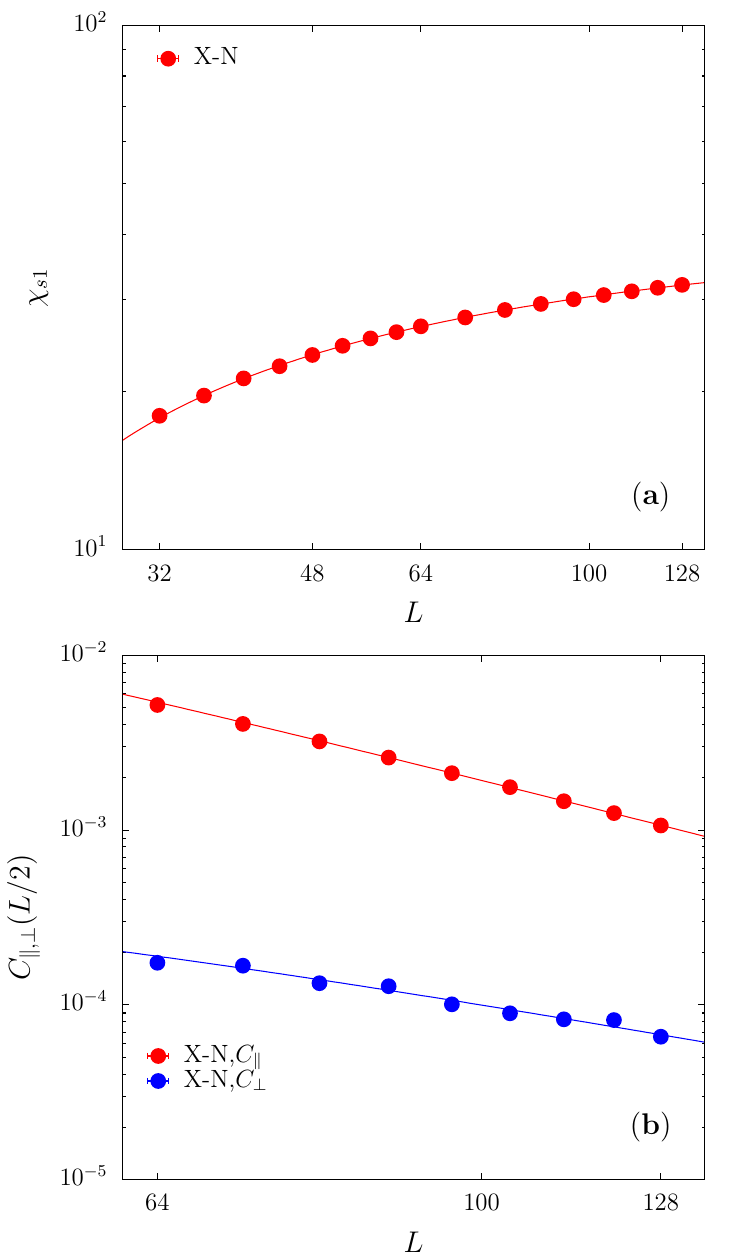}
	\caption{Surface staggered magnetic susceptibility $\chi_{1s}$ (a) 
and the correlations $C_{\parallel}(L/2)$ and $C_{\perp}(L/2)$ (b) versus system size $L$ 
on the XN surface configurations. The plots are on a log-log scale.}
	\label{fig:xncritial}
\end{figure}

At the bulk critical point, we expect the surface to exhibit critical singularities purely induced by the bulk criticality, belonging to the ordinary class. 
Our numerical results confirm this expectation.

The numerical result of $\chi_{1s}$ as a function of size $L$ is graphed in  Fig. \ref{fig:xncritial}(a), and the results 
of $C_{\parallel}(L/2)$ and $C_{\perp}(L/2)$ as functions of $L$ are plotted in Fig. \ref{fig:xncritial}(b).
We fit the data of  $C_{\parallel}(L/2)$ and $C_{\perp}(L/2)$ according to Eqs. (\ref{sucs1}) and (\ref{sucs2}) and 
find statistically sound estimates of $\eta_{\parallel}=1.57(18)$ and $\eta_{\perp}=0.82(8)$.
The finite-size scaling form Eq. (\ref{suchis}) supplemented with a constant $c$ as non-singular 
contribution is used to fit the data of $\chi_{1s}$. The estimate of $y_{h1}$ is 0.68(7).  
The obtained surface exponents satisfy the scaling relations in Eqs. (\ref{scalings1}) and  agree well with the surface universal class of the ordinary transition associated with the 
3D Ising universality class \cite{ Deng2005}.

\subsection{Extraordinary SCBs on the Y surface}

We then study the surface critical behaviors on the Y surface,  illustrated in Fig. \ref{Fig:model} (c). A 1D usual ladder behaves like a spin-1 chain\cite{diagonal2000}, and the ground state is described by the Affleck-Kennedy-Lieb-Tasaki (AKLT) state \cite{affleck1988valence,liu2022bulk,liu2023probing,liu2023measuring}. With free boundaries, each end of a ladder has one spin-1/2 spinon located \cite{Abanov2017}.

\begin{figure}[htb]
	\centering
	\includegraphics[width=0.46\textwidth]{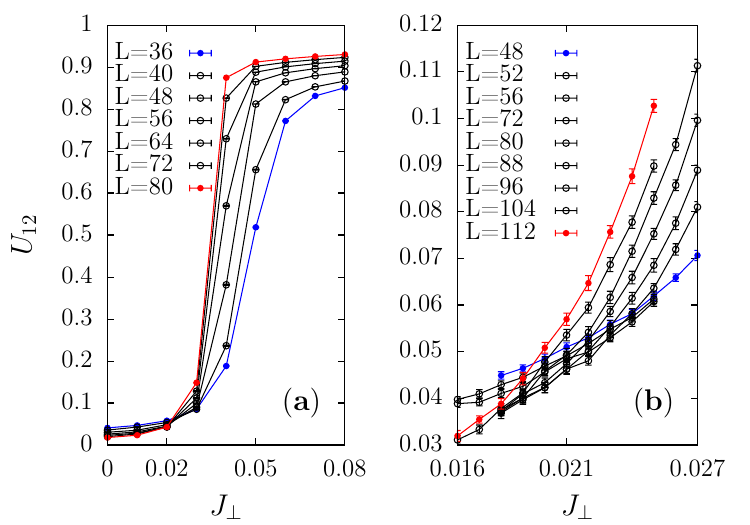}
	\caption{(a) Surface Binder cumulant $U_{12}$ versus $J_{\perp}$ 
	for different system sizes. Error bars are much smaller than the symbols. 
 (b) shows more detailed data around the crossing point. }
	\label{fig:ybinder}
\end{figure}

\begin{figure}[htb]
	\centering
	\includegraphics[width=0.46\textwidth]{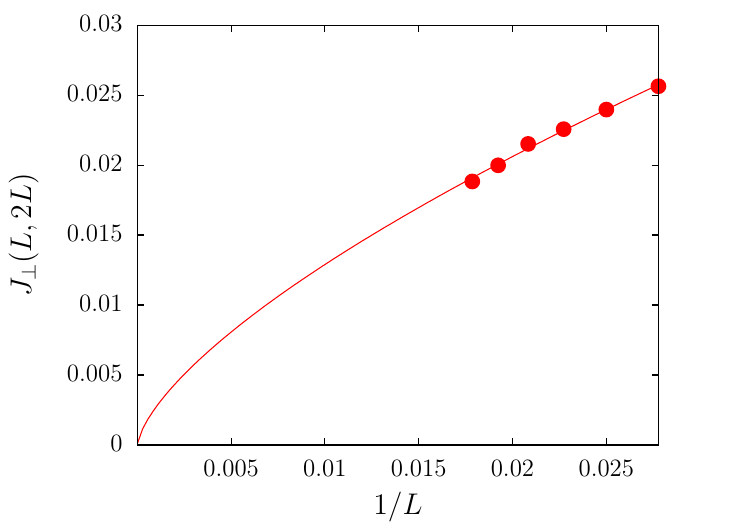}
	\caption{$J_{\perp}(L,2L)$  versus $1/L$. The standard $(L,2L)$ crossing analysis applied to the crossing points $J_{\perp }(L, 2L)$ of $U_{12}(J_\perp, L)$.
 Fitting the crossings with  $J_{\perp}(L,2L) = bL^{-p}$, we obtain statistically sound estimates of $b=0.61(8)$ and $p=0.86(3)$, suggesting convergence to 0. 
}
	\label{fig:cross}
\end{figure}

When the ladders are coupled weakly($J_\perp < 0.040006$), the system stays in the Q1DH phase, and the spin-1/2 spinons are coupled 
through Ising couplings, forming an effective AF Ising chain. This contrasts with the AF Heisenberg chain, 
which is a $(1+1)$D $SU(2)_1$ CFT. Therefore, the Y surface should exhibit AF Ising order. 
To verify this, we calculate the Binder cumulant $U_{12}$ on the surface as functions of $J_{\perp}$ for different system sizes, as shown in Figure \ref{fig:ybinder}.
At first glance, it appears that the surface undergoes a phase transition at around $J_{\perp}\sim 0.02$.
We apply the standard $(L,2L)$ crossing analysis to estimate the ``critical point"  \cite {shao_science}.  The crossing points extrapolate to zero for $L\to \infty$, as illustrated in 
Fig. \ref{fig:cross}, indicating the  ``critical point" does not exist and the surface is ordered for all $J_{\perp}>0$.

Such surface property can be attributed to the bulk-edge correspondence of the SPT Q1DH state, which leads to the presence of nontrivial surface states that are either gapless or degenerate and cannot be eliminated if the protecting symmetries are preserved \cite{XiaoGang2009, Frank2010, XiaoGang2012}.

\begin{figure}[h!]
	\centering
	\includegraphics[width=0.4\textwidth]{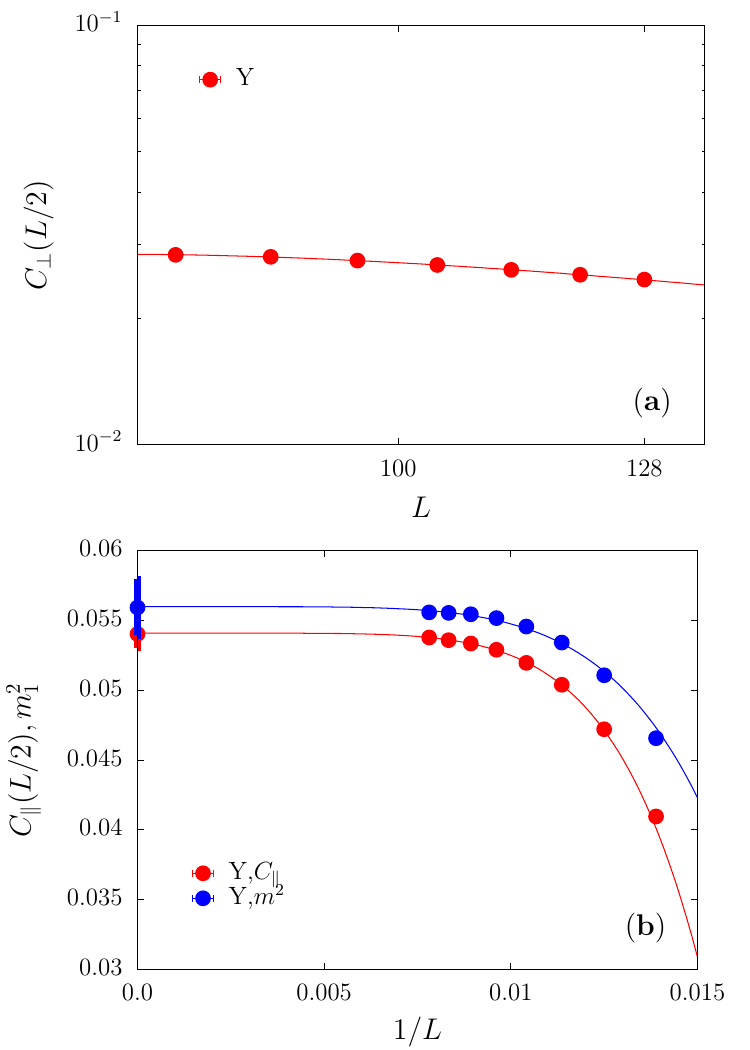}
	\caption{ (a) The correlations  $C_{\perp}(L/2)$ versus system size $L$ on a log-log scale at the bulk critical point and (b) Squared surface ferromagnetic magnetization $m_1^2$  and the correlations $C_{\parallel}(L/2)$ versus $1/L$. }
	\label{fig:ycritial}
\end{figure}

When the ordered surface is coupled to the critical bulk,  extraordinary SCBs of the 3D Ising UC are anticipated. 

The numerical results of $C_{\parallel}(L/2)$ and $m^2_{1s}(L)$ as functions of size $L$ at the bulk critical point are graphed in 
Fig. \ref{fig:ycritial}(b). Fitting these
data according to Eqs. (\ref{sum1s}) and (\ref{sucs1}), we obtain $C_\parallel=0.054(1)$ and $m^2_{1s}=0.055(2)$, which are consistent within error bars. This indicates the existence of a long-range order on the surface. 
However, it is difficult to obtain a meaningful estimate of $\eta_\parallel$ and $y_{h1}$. The reason is that $\eta_\parallel$ is big,  while $y_{h1}$ is very small, making it difficult to fit Eqs. (\ref{sum1s}) and (\ref{sucs1}) to the data. 

The $C_{\perp}(L/2)$ as a function of $L$ are plotted in Fig. \ref{fig:ycritial}(a). The finite-size scaling form in Eq. (\ref{sucs2}) is used in the fitting to the data $C_{\perp}(L/2)$. Our estimate of the exponent $\eta_{\perp}=-0.19(3)$ with the leading correction $1/L$ included.
The surface singularities are induced by the bulk criticality.
These results show that the surface indeed shows extraordinary SCBs at the bulk critical point.


\subsection{Unconventional surface transition and SCBs on the XD surface}

The XD surface is exposed by cutting the strong FM bonds $J_1$, as shown in Fig. \ref{Fig:model} (d). 
The surface is a spin-1/2 AF Heisenberg chain formed by dangling spins.  
Therefore, it is gapless in the Q1DH phase with the couplings to the bulk weak. 
This is verified by the power-law decay in $C_\parallel(L/2)$.

\begin{figure}[htb]
        \centering
        \includegraphics[width=0.46\textwidth]{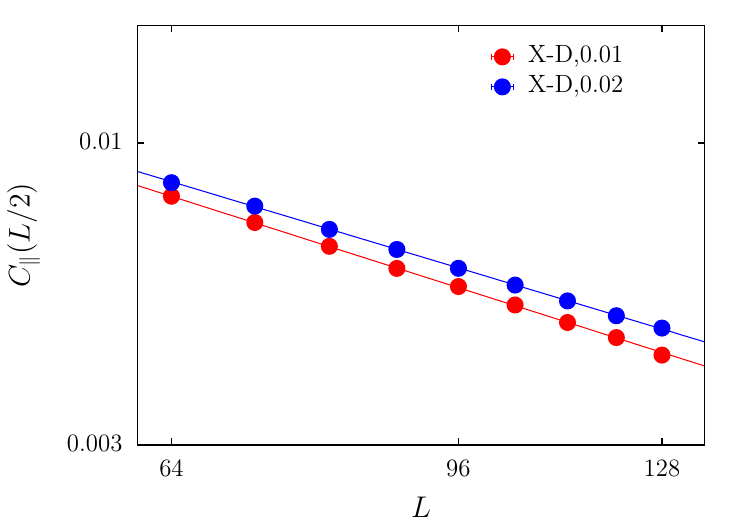}
        \caption{Surface correlation $C_{\parallel}(L/2)$ vs. system size $L$ in the Q1DH phase ($J_{\perp}=0.01,0.02$). The plot is set on a log-log scale. Power-log decaying with $L$ is observed, meaning the surface state is gapless.}
        \label{fig:xdgap}
\end{figure}

Figure \ref{fig:xdgap} shows $C_\parallel(L/2)$ at $ J_\perp=0.01$ and $ J_\perp=0.02$ sitting in the Q1DH phase. We see that $C_\parallel(L/2)$ decays with system size $L$ in a power law as follows:
\begin{equation}
    C_\parallel(L/2) \sim  L^{-p}.
    \label{gapless_surface}
\end{equation}
We find $p=0.90(1)$ for the $J_\perp=0.01$ and $p=0.86(1)$ for the $J_\perp=0.02$,
meaning that  surface state is gapless.

According to Mermin-Wagner theorem\cite{Mermin1966}, a gapless AF Heisenberg chain, which is a (1+1)D $SU(2)_1$ CFT, can not spontaneously break the $SU(2)$ symmetry or its $U(1)$ subgroup. However, if the chain is coupled to bulk critical fluctuations towards the (2+1)D $Z_2$ UC, which still preserves the $U(1)$ symmetry,  it should develop an Ising order before the bulk hits criticality, which is similar to that discussed in Ref.\cite{Jian2021}. Such a transition is not a (1+1)D Ising transition but in the KT-like UC \cite{Haldane}.

\begin{figure*}[htb]
	\centering
	\includegraphics[width=0.99\textwidth]{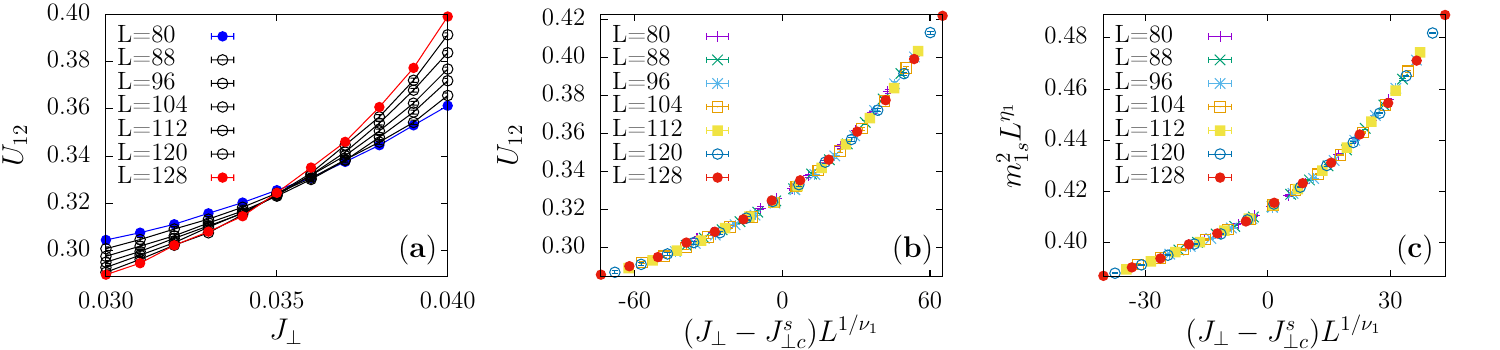}
	\caption{Surface transition before the bulk orders on the XD surface. (a) Surface Binder cumulant $U_{12}$ versus $J_{\perp}$ 
	for different system sizes. (b) Scaling with $\nu_1=0.86(5)$ and $J_{\perp c}^s=0.03478(8)$ of $U_{12}$. (c) $m_{1s}^2$ scaled using $\nu_1=0.88(2)$, $J_{\perp c}^s=0.03478$ and $\eta_1=0.6659(7)$. Error bars are much smaller than the symbols.}
	\label{fig:surfacedc}
\end{figure*}

Indeed, we find a surface transition before the bulk orders by studying the finite-size scaling of the surface Binder cumulant $U_{12}$. Figure \ref{fig:surfacedc} (a) plots $U_{12}$ as functions of $J_{\perp}$ for different system sizes. Since $U_{12}$ is dimensionless at a QCP, the crossings of curves $J_\perp \approx 0.035$ for different sizes roughly indicate a transition point. For coupling larger than 0.035, we see the Ising order showing.
The scaling form Eq.(\ref{scaling}) is used 
to extract the critical coupling $J^s_{\perp c}$ and the surface correlation length exponent $\nu_1$
for $A=U_{12}$ and $\kappa=0$. 
By expanding the scaling function into polynomials near the surface transition point, we find $J^s_{\perp c}=0.03478(8)<J_{\perp c}$ and $\nu_1=0.86(5)$ by fitting the polynomials to $U_{12}(J_\perp, L)$. Details of the fitting are presented in Appendix \ref{sec:appd}.
Perfect data collapsing is seen in Fig. \ref{fig:surfacedc} (b) using the obtained $J^s_{\perp c}$ and  $\nu_1$. 

We then calculate the squared surface magnetization, $m_{1s}^2 = \langle (m_{1s}^z)^2\rangle$,
to extract the anomalous exponent $\eta_1$ 
according to the scaling form Eq.~(\ref{scaling}) with $A=m_{1s}^2$ and $\kappa=-\eta_1$. 
Setting $J_{\perp c}^s=0.03478$, we obtain $\nu_1=0.88(2)$ and $\eta_1=0.6659(7)$. Within the error bar, $\nu_1=0.88(2)$ is the same as $\nu_1=0.86(5)$ obtained from $U_{12}$. 
Details of the fitting are available in Appendix \ref{sec:appd}.
Perfect data collapsing is seen in Fig. \ref{fig:surfacedc} (c) using the obtained $J^s_{\perp c}$, $\nu_1$ and $\eta_1$. 

Our results support that the surface transition is not a conventional (1+1)D Ising transition, but also
do not agree with the KT-like phase transition. 

\begin{figure}[htb]
	\centering
	\includegraphics[width=0.49\textwidth]{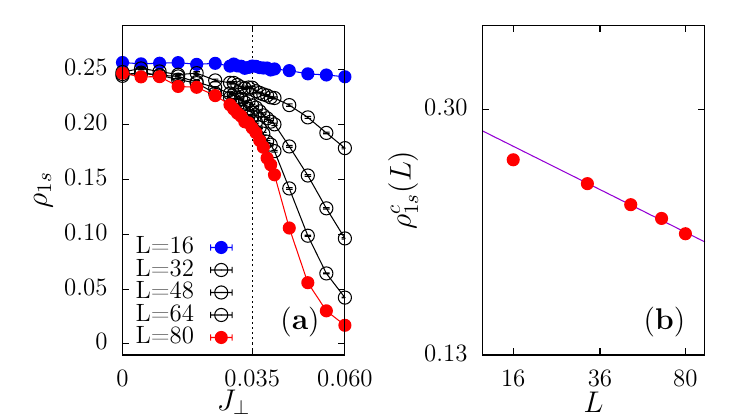}
	 \caption{(a)The  surface spin stiffness $\rho_{1s}(L)$ versus $J_{\perp}$ 
	 for different system sizes. The vertical dashed line corresponds to the critical point $J_{\perp c}^s$ obtained by binder cumulant $U_{12}$.  (b) The plot on a log-log scale and the line is the fit according to  $\rho^c_{1s}(L) \sim  L^{-p}$.}
	\label{fig:stiff}
\end{figure}

To further characterize the transition, we study the surface spin stiffness $\rho_{1s}$ which is defined as \cite{Pollock1987,SandvikAIP}
\be
\rho_{1s}=\langle(N^+_x-N^-_x)^2\rangle/(\beta L),
 \ee
where $N^+_x$ and $N^-_x$ denote the total number of operators 
transporting spin in the positive and negative $x$ direction on the surface. 
Figure \ref{fig:stiff}(a) shows $\rho_{1s}(L)$ as functions of $J_\perp$ for different size $L$.

Suppose the transition is of KT-like, $\rho_{1s}(L)$ at $J_{\perp c}^s$ is expected to scale as 
\cite{Weber1988,SandvikAIP}
\be
\rho^c_{1s}(L)=\rho^c_{1s} (1+\frac{1}{2 \ln(L)+c}),
\label{rho1}
\ee
 where $c$ is a nonuniversal constant, $\rho^c_{1s}$ is a finite spin stiffness at the transition point in the thermodynamic limit. 
 However, we find that this scaling form does not fit our finite-size data of $\rho^c_{1s}(L)$.  
 Instead, we find $\rho^c_{1s}(L)$ converges to zero in a power law $\rho^c_{1s}(L) \sim  L^{-p}$ with $p=0.18(1)$, as shown in Fig. \ref{fig:stiff}(b). 
 Details of the fitting are presented in Appendix \ref{sec:appe}. 

In previous research for quantum systems with O(3) symmetry,  nonordinary SCBs are expected at surfaces with gapless surface mode 
due to the merging of gapless surface mode with the 3D O(3) bulk critical modes. Such nonordinary SCBs can be multicritical special SCBs 
\cite{Zhang2017,Weber2018,Ding2018,Zhu2021} or extraordinary SCBs with surface ordered \cite{Wang2023}.
For the current model, the bulk undergoes the 3D $Z_2$ phase transition. 
When the system hits the bulk criticality, the Ising ordered surface is coupled to the (2+1)D Ising bulk critical mode. 
We thus expect extraordinary SCBs in the (2+1)D Ising UC, which should be the same as that on the 
Y surface.
However, we obtain different critical behaviors from the Y surface, in particular, we find $\eta_\perp = -0.68(8)$. 

\begin{figure}[t]
	\centering
	\includegraphics[width=0.46\textwidth]{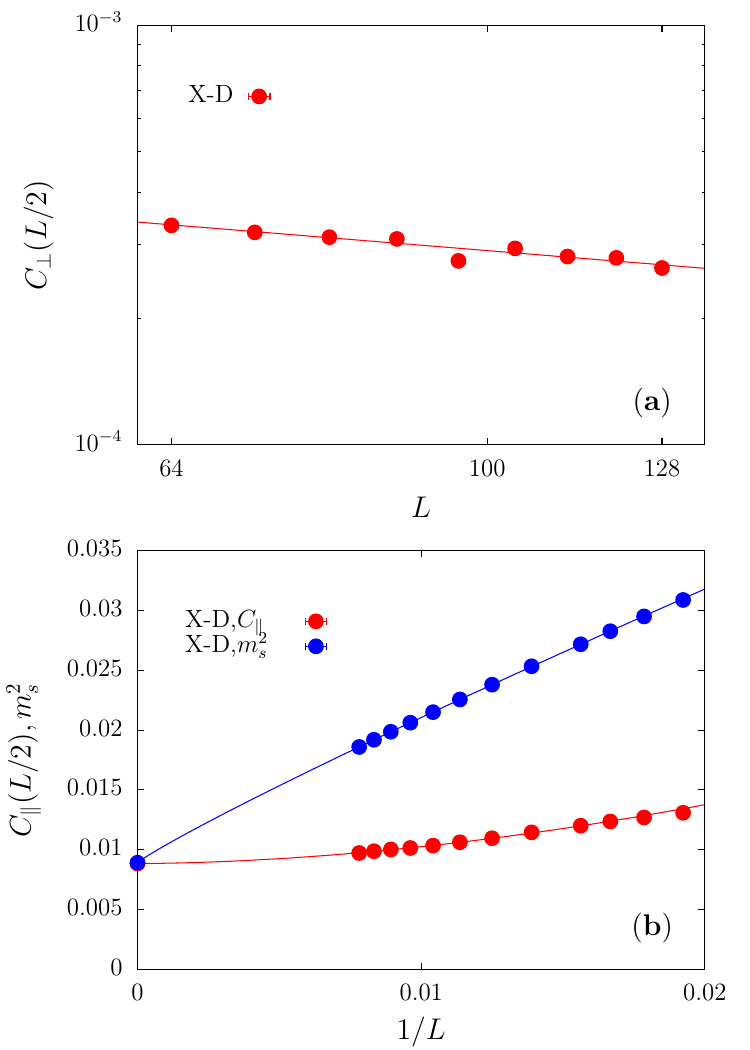}
	\caption{(a) The correlations  $C_{\perp}(L/2)$ versus system size $L$ on a log-log scale at the bulk critical point and (b) Squared surface antiferromagnetic magnetization $m_1^2$  and the correlations $C_{\parallel}(L/2)$ versus $1/L$. }
	\label{fig:xdcritical}
\end{figure}

The numerical results of $C_{\parallel}(L/2)$ and $m^2_{1s}(L)$ as functions of size $L$ at the bulk critical point are graphed in 
Fig. \ref{fig:xdcritical}(b). Fitting these
data according to Eqs. (\ref{sucs1}) and (\ref{sum1s}), we obtain $C_\parallel=0.0089(4)$ and $m^2_1=0.0088(2)$, which are consistent within error bars. 
This indicates the existence of a long-range order on the surface. However, it is also difficult to obtain a meaningful estimate of $\eta_\parallel$ and $y_{h1}$. 
The $C_{\perp}(L/2)$ as a function of $L$ are plotted in Fig. \ref{fig:xdcritical}(a). The finite-size scaling form in Eq. (\ref{sucs2}) is used to fit the data 
of $C_{\perp}(L/2)$. Our final estimate of the exponent $\eta_{\perp}$ is $\eta_{\perp}=-0.68(8)$. 

\section{Discussion and Conclusion}
\label{sec:conclsn}
In this paper, we designed a quantum spin-1/2 model hosting a quantum phase transition in the (2+1)D Ising universality class.
We have shown that the model exhibits rich surface properties, including ordinary SCBs on the XN surface,  extraordinary SCBs on the 
Y surface, and in particular exotic surface phase transition and unconventional
extraordinary SCBs on the XD surface.

The exotic surface transition refers to the surface ordering before the bulk as the Ising couplings increase,
which is different from the surface critical behaviors associated with the bulk critical point, which occur at the bulk critical point.
This surface transition seems to have critical exponents different from those of the (1+1)D Ising criticality and the KT transition 
predicted in previously theoretical work.
In a perturbation theory, one can easily obtain an effective $S^z_iS^z_j$ interaction at the 
surface from couplings to the bulk fluctuations, leading to an effective (1+1)D XXZ-type
chain with U(1) symmetry, which should induce a KT-like phase transition 
here \cite{yan2018interacting}. 
On the other hand, if the Ising couplings to the bulk are relevant, the surface may order before the 
bulk, but the transition should be in the (1+1)D Ising UC, which would lead to the extraordinary SCBs in the 3D Ising UC,  similar to the extraordinary SCBs on the Y surface.
However, our numerical results show conflict with these expectations. Therefore,
the unconventional surface transition and extraordinary SCBs obtained here surpass our current understanding, which may hint at some new scenario of surface critical behavior that desires further investigations both numerically and theoretically.

\begin{acknowledgments}
We acknowledge the helpful discussion with Shuai Yin and Chao-Ming Jian.
Z.W. and Z.Y. thank the support from the start-up fund of the Westlake University. W.G. thanks the support from the National Natural Science Foundation of China under Grant No.~12175015. Y.C.W. acknowledges the support from Zhejiang Provincial Natural Science Foundation of China (Grant No. LZ23A040003), and the support from the High-Performance Computing Centre of Hangzhou International Innovation Institute of Beihang University. The authors also acknowledge Beijing PARATERA Tech Co., Ltd. and the HPC centre of Westlake University for providing HPC resources.  SQN is supported by Direct Grant
No.4053578 from The Chinese University of Hong Kong, and funding from Hong Kong’s Research Grants Council (GRF No.14306420). Z.W. is supported by the China Postdoctoral Science Foundation under Grants No.2024M752898.
\end{acknowledgments}

\clearpage
\appendix

\section{Bulk critical properties for systems with aspect ratio $ R=4$ }
\label{sec:r4}

\begin{figure}[h]
\centering
\includegraphics[width=0.46\textwidth]{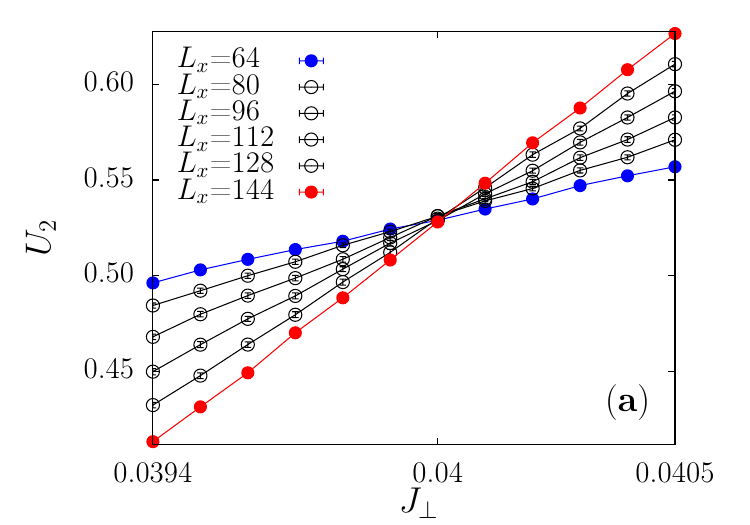}
\caption{ Binder cumulant $U_{2}$ versus $J_{\perp}$ for different system sizes with $R=4$. }
\label{fig:r4binder}
\end{figure}

\begin{figure}[h!]
\centering
\includegraphics[width=0.46\textwidth]{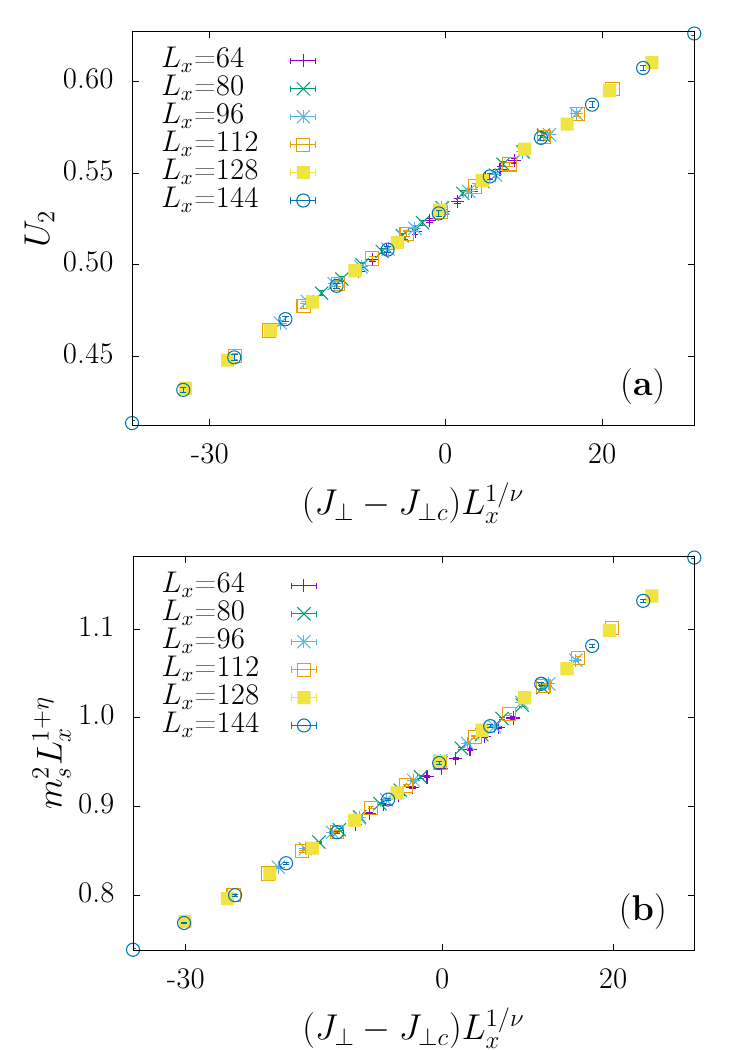}
\caption{ Critical properties of the bulk phase transition for systems with $R=4$. (a) Data collapse of $U_2$ using $\nu=0.63$ and $J_{\perp c}=0.04001$. (b) Data collapse of $m_s^2$ using $\nu=0.63$, $J_{\perp c}=0.04001$, and $\eta=0.033$. Error bars are much smaller than the symbols.}
\label{fig:r4collapse}
\end{figure}

We present here the simulation results for systems with an aspect ratio $ R=L_x/L_y=4$ and inverse temperature $\beta= L_x$, with PBCs 
applied along both $x$ and $y$ lattice directions to study the critical properties near the critical point.  

Figure \ref{fig:r4binder} shows $U_2$ as a function of $J_{\perp}$ for different system sizes. 
The system with $L_x=64$ has the same number of sites as the system with $L=32$ and $R=1$. 
Comparing to Fig. \ref{fig:binder}, we see the shifts of the crossings for two neighboring sizes become much narrower as sizes increase, 
indicating that the correction to scaling due to annisotropy is greatly reduced by the aspect ratio.
We fit the finite-size scaling formula Eq.~(\ref{scaling}) to the data of $L_x \ge 64$
to extract  the critical coupling $J_{\perp c}$ and the correlation length exponent $\nu$, where $f(x)$ is a scaling function. 
For $A=U_2$ which is dimensionless,  $\kappa=0$. Expanding $f(x)$ to polynomials near the transition point, we obtain 
$J_{\perp c}=0.04001(1)$ and  $\nu=0.63(2)$ by fitting the polynomials to the finite-size data. 
Figure \ref{fig:r4collapse} (a) shows a perfect data collapsing of $U_2$ using the obtained $J_{\perp c}$ and $\nu$. 

Next, the squared magnetization $m_s^2$ is analyzed to extract the exponent $\eta$. The scaling of $m_s^2$ follows the form given in 
Eq.~(\ref{scaling}), with $\kappa=-(1+\eta)$ and $f(x)$ being a scaling function, ignoring the correction to scaling $L^{-\omega}$. 
Using the obtained  $J_{\perp c}$ and $\nu$, 
we obtain $\eta=0.033(2)$ by expanding $f(x)$ to polynomials and fitting it to finite-size data of $m_s^2(J_\perp,L)$. Perfect data 
collapse using obtained critical properties is shown in Fig. \ref{fig:r4collapse} (b).  

Here, the obtained $J_{\perp c}=0.04001(1)$, $\nu=0.63(2)$ 
and $\eta=0.033(2)$ from systems with $R=4$ are consistent with $J_{\perp c}=0.040006(3)$,  $\nu=0.64(1)$ and $\eta=0.036(3)$  
obtained from systems with $R=1$ within the error bars. The larger error bars for the $R=4$ results are due to the fact 
that the largest system size used are much smaller than that used for systems with $R=1$.

\section{Unconventional surface transition and extraordinary surface critical behaviors on the XD surface}
\label{sec:appd}



In the main text, we show that the boundary will go through a transition from the gapless state to an Ising phase before the bulk hits criticality. 


Here, we first present the details fitting about surface Binder cumulant $U_{12}$. We adopt finite-size scaling as derived from 
Eq. (\ref{scaling}) 
\begin{equation}
\begin{split}
U_{12}(J_\perp, L) &=f[(J_{\perp}-J^s_{\perp c})L^{1/\nu_1}, L^{-\omega}]  \\
&=U_{0}+\sum\limits_{k=1}^{k=6}a_{k}(J_{\perp}-J^s_{\perp c})^{k}L^{k(1/\nu_1)}\\
&+b(J_{\perp}-J_{\perp c}^s)L^{1/\nu_1-\omega}+cL^{-\omega},
\end{split}
\end{equation}
to extract  the critical coupling $J^s_{\perp c}$ and the correlation length exponent $\nu_1$.  Table \ref{extr} shows fitting results without including effective correction to scaling $L^{-\omega}$,   and Table \ref{extrc} shows results with effective correction to scaling $L^{-\omega}$  included.

\begin{ruledtabular}
\begin{table}[!h]
\caption{ Fitting $U_{12}$	without correction-term.  Parameters to be used in fitting 
 from $J_{\perp}=0.029$ to $J_{\perp}=0.038$. $L_{min}$ is the minimum system size used in fitting. Reduced $\chi^2$ (R-$\chi^2$) and p-value of $\chi^2$ (P-$\chi^2$) are also listed. }
\begin{tabular}{l c c c   }
 	   	   $L_{min}$ 	  & $J_{\perp c}^s$ 	  & $\nu$ 	&R/P-$\chi^2$ \\
 	   	   	
\hline
	$80$			& 0.035375(8)   	&0.82(5)   &3.22/0.000029  \\   
	$88$			& 0.03507(8)  	&0.83(5)   &1.92/0.00019 \\
    $96$			& 0.03478(8)   	&0.86(5)   &1.22/0.17 \\
	$104$			& 0.0347(1)   	&0.89(6)  &1.37/0.09 \\
\end{tabular}
\label{extr}
\end{table}
\end{ruledtabular}

\begin{ruledtabular}
\begin{table}[!h]
\caption{ Fitting $U_{12}$	with correction-term. Parameters to be used in fitting 
 from $J_{\perp}=0.029$ to $J_{\perp}=0.038$. $L_{min}$ is the minimum system size used in fitting. Reduced $\chi^2$ (R-$\chi^2$) and p-value of $\chi^2$ (P-$\chi^2$) are also listed.}
\begin{tabular}{l c c c   }
 	   	   $L_{min}$ 	  & $J_{\perp c}^s$ 	  & $\nu$ 	&R/P-$\chi^2$ \\
 	   	   	
\hline
	$80$			& 0.030(5)   	&1.5(4)   &0.99/0.49  \\   
	$88$			& 0.033(2)  	&0.82(29)   &1.57/0.01 \\
    $96$			& 0.034(2)   	&0.96(29)   &1.23/0.17 \\
	$104$			& 0.036(2)   	&0.84(42)  &1.62/0.03 \\
\end{tabular}
\label{extrc}
\end{table}
\end{ruledtabular}

For the squared surface magnetization $m_{1}^2$,  we adopt the following finite-size scaling formula
\begin{equation}
\begin{split}
m_{1}^2(J_\perp, L) &=L^{-\eta_1}f[(J_{\perp}-J_{\perp c}^s)L^{1/\nu_1}, L^{-\omega}] \\
&=L^{-\eta_1} (m_{0}+\sum\limits_{k=1}^{k=3}a_{k}(J_{\perp}-J_{\perp c}^s)^{k}L^{k(1/\nu_1)}\\
&+b(J_{\perp}-J_{\perp c}^s)L^{1/\nu_1-1}
+cL^{-1}),
\end{split}
\end{equation}
to extract the correlation length exponent $\nu_1$ and the exponent $\eta_1$.  We here take the effective correction to scaling exponent $\omega=1$. Without correction-term (Table \ref{extm}) and with correction-term (Table \ref{extmc}) are obtained.

\begin{ruledtabular}
\begin{table}[!h]
\caption{ Fitting $m_{1}^2$	without correction-term.  Parameters to be used in fitting 
 from $J_{\perp}=0.03$ to $J_{\perp}=0.04$. $L_{min}$ is the minimum system size used in fitting. Reduced $\chi^2$ (R-$\chi^2$) and p-value of $\chi^2$ (P-$\chi^2$) are also listed.}
\begin{tabular}{l c c c   }
 	   	   $L_{min}$ 	  & $\nu$ 	  & $\eta$ 	&R/P-$\chi^2$ \\
 	   	   	
\hline
	$80$			& 0.88(2)   	&0.6659(7)   &1.58/0.002 \\   
	$88$			& 0.89(2)  	&0.6638(9)   &1.45/0.02 \\
    $96$			& 0.90(3)   	&0.662(2)   &1.41/0.03 \\
	$104$			& 0.90(3)   	&0.662(2)  &1.58/0.02 \\
\end{tabular}
\label{extm}
\end{table}
\end{ruledtabular}

\begin{ruledtabular}
\begin{table}[!h]
\caption{ Fitting $m_{1}^2$	with correction-term.  Parameters to be used in fitting 
 from $J_{\perp}=0.03$ to $J_{\perp}=0.04$. $L_{min}$ is the minimum system size used in fitting. Reduced $\chi^2$ (R-$\chi^2$) and p-value of $\chi^2$ (P-$\chi^2$) are also listed.}
\begin{tabular}{l c c c   }
 	   	   $L_{min}$ 	 & $\nu$ 	  & $\eta$ 	&R/P-$\chi^2$ \\
 	   	   	
\hline
	$80$			& 0.97(3)   	&0.54(3)   &1.18/0.14  \\   
	$88$			& 0.97(3)  	&0.55(5)   &1.30/0.06 \\
    $96$			& 0.98(7)   	&0.57(8)   &1.40/0.04 \\
	$104$			& 0.96(11)   	&0.57(14)  &1.59/0.02 \\
\end{tabular}
\label{extmc}
\end{table}
\end{ruledtabular}


\section{Another evidence for negating the KT phase transition}
\label{sec:appe}
\begin{figure}[htb]
	\centering
	\includegraphics[width=0.46\textwidth]{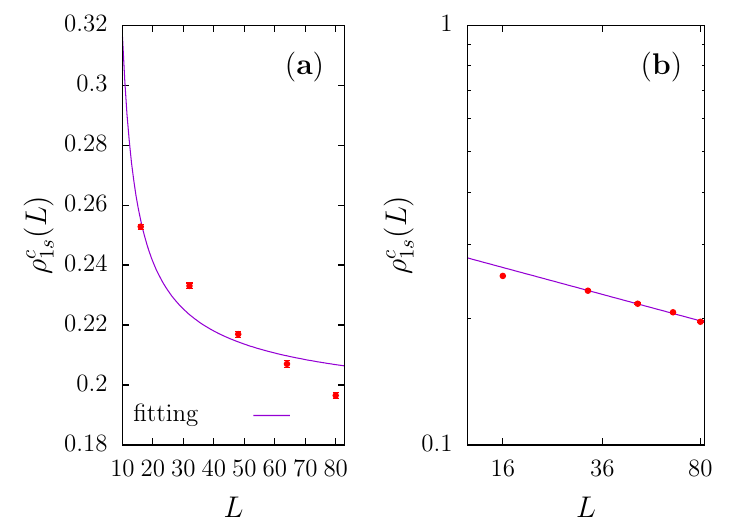}
	\caption{  The spin stiffness at the critical point, $ \rho^c_{1s}(L)$,  versus system size $L$. 	(a) The plot on a conventional scale and the line is the fit according to Eq.~(\ref{rho1}). (b) The plot on a log-log scale and the line is the fit according to Eq.~(\ref{rho2}).}
	\label{fig:rho1}
\end{figure}

At the KT phase transition point, the
spin stiffness  is expected flows as a function of L as Eq. (\ref{rho1}) \cite{Weber1988,SandvikAIP}.
As shown in Fig. \ref{fig:rho1} (a), we find that the $\rho^c_{1s}(L)$ can not be fitted by Eq.~(\ref{rho1}). We also try to fit the $\rho^c_{1s}(L )$ with a power-law,
 \begin{equation}
    \rho^c_{1s}(L) \sim  L^{-p}.
    \label{rho2}
\end{equation}
and find statistically sound estimate  $p=0.18(1)$ (see Fig. \ref{fig:rho1} (b) ). All the above results show that the phase transition is not a KT phase transition.

\end{document}